# Dynamic properties of bacterial pili measured by optical tweezers

Erik Fällman[a], Magnus Andersson[a], Staffan Schedin[b], Jana Jass[c], Bernt Eric Uhlin[d], and Ove Axner[a]

[a]Dept. of Physics, Umeå University, SE-901 87 Umeå, Sweden.
[b]Dept. of Applied Physics and Electronics, Umeå University, SE-901 87 Umeå, Sweden.
[c]The Lawson Health Research Institute, 268 Grosvenor St., London, Ontario, Canada, N6A 4V2.
[d]Dept. of Molecular Biology, Umeå University, SE-901 87 Umeå, Sweden.

**ABSTRACT**

The ability of uropathogenic *Escherichia coli* (UPEC) to cause urinary tract infections is dependent on their ability to colonize the uroepithelium. Infecting bacteria ascend the urethra to the bladder and then kidneys by attaching to the uroepithelial cells via the differential expression of adhesins. P pili are associated with pyelonephritis, the more severe infection of the kidneys. In order to find means to treat pyelonephritis, it is therefore of interest to investigate the properties P pili. The mechanical behavior of individual P pili of uropathogenic *Escherichia coli* has recently been investigated using optical tweezers. P pili, whose main part constitutes the PapA rod, composed of ~1000 PapA subunits in a helical arrangement, are distributed over the bacterial surface and mediate adhesion to host cells. We have earlier studied P pili regarding its stretching/elongation properties where we have found and characterized three different elongation regions, of which one constitute an unfolding of the quaternary (helical) structure of the PapA rod. It was shown that this unfolding takes place at an elongation independent force of 27 ± 2 pN. We have also recently performed studies on its folding properties and shown that the unfolding/folding of the PapA rod is completely reversible. Here we present a study of the dynamical properties of the PapA rod. We show, among other things, that the unfolding force increases and that the folding force decreases with the speed of unfolding and folding respectively. Moreover, the PapA rod can be folded-unfolded a significant number of times without loosing its characteristics, a phenomenon that is believed to be important for the bacterium to keep close contact to the host tissue and consequently helps the bacterium to colonize the host tissue.

## 1. INTRODUCTION

Adhesion of bacteria to host tissue is an essential step in the progression of an infection. Many Gram-negative bacteria express pili, which are responsible for mediating adhesion and maintaining bacteria-host contact during the initial stages of an infection by many Gram-negative bacteria. Uropathogenic *Escherichia coli*, which are implicated in 75-80% of uncomplicated urinary tract infections and in severe pyelonephritis, express a variety of different fimbrial adhesins of which P pili predominantly are expressed by isolates from the upper urinary tract [1, 2]. Thus P pili, expressed by approximately 90% of the *E. coli* strains that cause pyelonephritis (upper urinary tract and kidney infections), constitute an important virulence factor [3].

Individual P pili have previously been characterized with respect to their static structure [4-13] (sizes and three-dimensional shape) as well as their elongation properties [14-16]. It has been found, for example, that they are assembled from six different structural proteins [13]. As is schematically illustrated in Panel B, *Fig. 9*, the main part of a P pilus constitutes a μm-long helical rod [6], the PapA rod, approximately 6.8 nm in diameter [7], with a short (15 nm long) and thin (2 – 3 nm in diameter) flexible fibrillum tip [12]. The PapA rod, which is composed of about a thousand 16.5kDa PapA subunits in a right-handed helical arrangement with 3.28 subunits per turn [7-10], is anchored to the membrane by PapH [5] and connected to the fibrillum by a PapK adaptor. The fibrillum, in turn, is a flexible structure composed of PapE [4], PapF [4] and the PapG adhesin [9, 11-13]. Since the major part of a P pilus is composed of PapA units, most of the mechanical (physical) properties of P pili are governed by the properties of the PapA rod.

We have by the use of an optical tweezers based force measurement system assessed various physical properties of individual P pilus. Below we describe the optical tweezers instrumentation that has been developed in our laboratory, the methodology used to perform accurate force measurements on pili structures, as well as some typical results.

## 2. INSTRUMENT DESIGN
### 2.1 The force measurement system

The system is constructed around an inverted microscope (Olympus IX71) equipped with a high numerical aperture oil-immersion objective as well as an oil immersion condenser. Components and modifications made to the microscope that is important for proper function of the force measurement system are pointed out in *Fig. 1*.

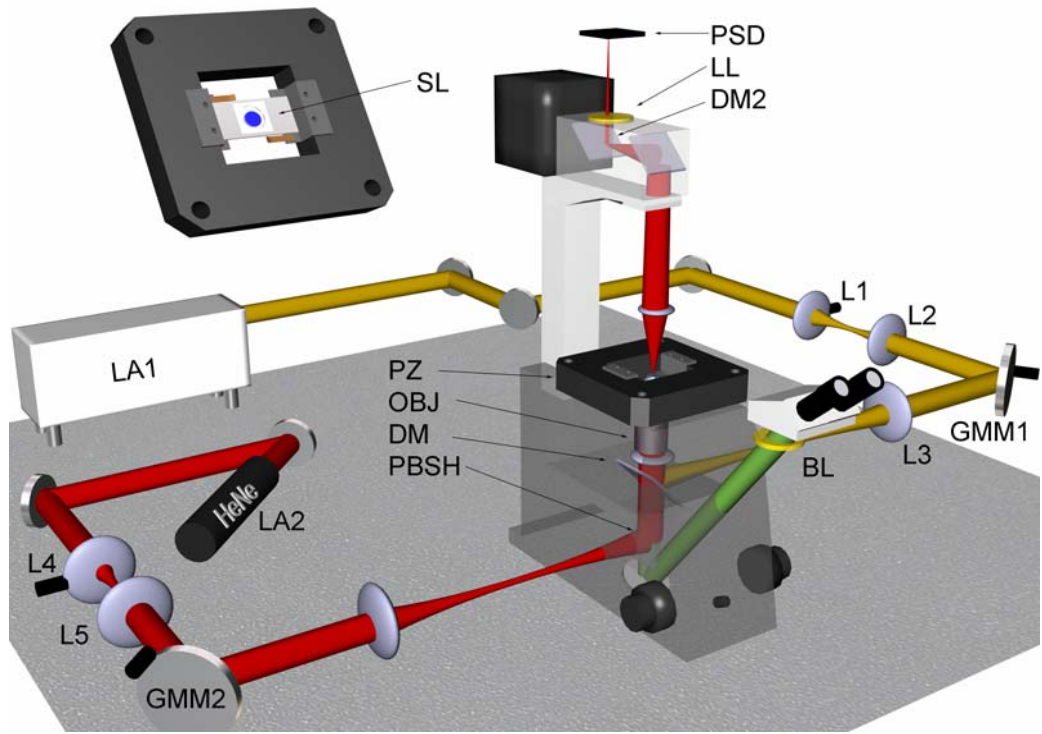

*Figure 1. Schematic illustration of the components in the force measurement system. The system consists of the following vital components: a cw Nd:YAG laser, LA1, for trapping; a HeNe laser, LA2, for probing the position of the object; a number of beam controlling lenses, L1-L6; a computer controlled gimbal mounted mirror, GMM1, for control of the trapping position; a dichroic mirror, DM, for introduction of the trapping light into the microscope; a polarizing beam splitter cube, PBSH, that merges the probe laser beam and the trapping laser beam; a block filter to prevent the laser light not to reach the eyepiece and the camera, BL; a high numerical aperture microscope objective, OBJ; a tailor made sample slide, SL; a piezo controlled table, PZ, for precise control of the position of the sample slide; an oil immersion condenser for collection of the probe laser light; a dichroic mirror, DM2, for deflection of the probe laser light onto the position sensitive detector; a laser line filter, LL, to block all light but the probe laser light from the detector; a x-y stage mounted sensitive devise, PSD, for probing the position of the probe laser light; a low noise preamplifier, and a measurement computer equipped with a data acquisition card and a GPIB card.*

The microscope is modified to merge the beams from the trapping laser and the probe laser with the light path of the microscope. The microscope is also equipped with additional filters to prevent laser light from reaching the eyes of the user. Moreover, a position sensitive detector has been added to the illumination arm for precise position detection of the trapped object. The trapping laser is feed into the microscope via the right side port of the microscope. The original 100% mirror of the right side port is replaced by a dichroic mirror, DM. The dichroic mirror is a plane plate coated with a high reflecting coating for the 1064 nm Nd:YAG-laser light on the surface facing the laser as well as an anti reflection coating for the Nd:YAG laser light on the opposite side (model #, HR1064HT633+337/45/BBAR RW33-25.4-3UV,SWP-L.-M.33x1, Laser components GmbH). Moreover the coating was chosen for high transmission of the 632 nm HeNe light as well as the visible spectrum. The probe laser is in turn conveyed into the microscope via the left side port where the original beam splitter cube is replaced with a polarizing beam splitter cube, PBSH (model no. PBSH-760-980-100; CVI laser Corporation, Albuquerque, MN, USA). The use of a polarizing beam splitter cube facilitates the merging of the probe laser beam with the light path of the microscope without loosing any laser power. Extra block filters are positioned before the eyepiece in the detection arm to prevent any laser light to reach the eyes of the user. A hot mirror is used for blocking the Nd:YAG (model no. hot mirror, 840; Omega Optical, Inc, USA)

The trapping laser is a cw Nd:YAG laser operating at 1064 nm (Millenia IR 10W Spectra Physics). For efficient trapping in the z-direction, the laser beam is expanded slightly larger than the entrance pupil of the microscope objective by two lenses in an afocal arrangement. A movable trap is constructed by an external lens-mirror arrangement such that the focus can be shifted in all three spatial directions. [17]. Movement along the horizontal (x-y) plane is achieved by tilting of a gimbal mounted mirror GMM1 (Model no. 07 MCD 515; Melles Griot, Täby, Sweden) positioned at an image plane of the entrance pupil of the microscope objective such that the laser beam pivots around the entrance aperture of the microscope objective when the mirror is tilted. To achieve high precise resolution of the movement of the trap in the x-direction GMM1 is in turn equipped with a piezo controlled mirror (Model S226, Physik Instrumente (PI) GmbH & Co. KG, Germany) Movement along the depth (z-direction) is achieved by changing the divergence of the beam by means of an afocal lens system where the lens L1 is stepper motor controlled (Model no. 07 TSC 517; Melles Griot, Täby, Sweden) The nominal position of L1 is at the sum of the focal lengths of L1 and L2 from the lens L2. The lens L2 is positioned fixed at a distance equal to its focal length from the gimbal mounted mirror, GMM1. This design results in that the degree of overfilling of the entrance aperture always remains the same. That is, the intensity at the focal point is constant regardless of the movement of the trap (i.e. the trapping power always stays the same). The lenses L1 and L2 are both antireflection-coated plan convex lenses (f = 60 mm, model no. LPX 125/083; Melles Griot, Täby, Sweden) positioned with the flat surface facing each other. Imaging of the entrance aperture onto the surface of the gimbal-mounted mirror is achieved by a second afocal lens system, which is produced by an external lens L3, (f = 250 mm, model no. LPX 282/083; Melles Griot, Täby, Sweden) and the fixed lens in the microscope beam path with a focal length of 180 mm.

The probe laser is a 7 mW polarized HeNe (model no. 1137P; Uniphase, Manteca, CA, USA). The position of the focus of the probe laser is controlled by an afocal system (L4, f = 100 mm and L5, f = 100 mm, model no. LPX 178/077; Melles Griot, Täby, Sweden) and a gimbal mounted mirror, GMM2. After the gimbal mounted mirror the probe laser light is fed through the lens L6, (f = 150 mm, model no. LPX 238/077; Melles Griot, Täby, Sweden) that equally with L3 forms an afocal system with the fixed lens in the microscope, i.e. the divergence of the light is maintained and the entrance pupil of the microscope objective is imaged onto the gimbal mounted mirror. To minimize the influence from the probe laser on the trapping efficiency the intensity of the probe laser is set to a fraction of that of the trapping laser by a neutral density filter. The power of the probe laser at the entrance pupil of the microscope objective was 0.45 mW. Both lasers are focused to a diffraction-limited spot by a high numerical aperture oil immersion microscope objective (model no. Olympus UPLFL100X/IR NA 1.30; Olympus) into the sample solution positioned in a piezo controlled sample chamber, *Fig. 2*.

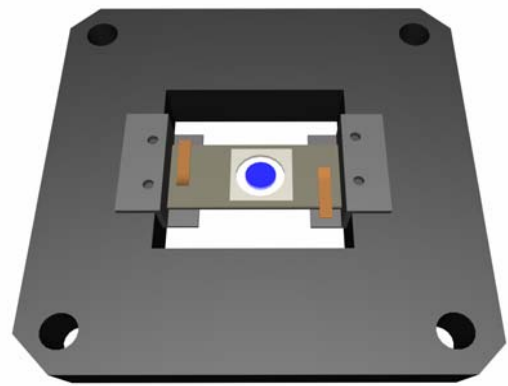

*Figure 2. A schematic illustration of the piezo controlled sample chamber: The sample slide is based on a 76x24x1.8 mm dural aluminium slide with a 18 mm circular cutout in the center and held by a tailor made piezo stage.*

The x-y position of the sample is precisely controlled via a tailor made sample slide holder, PZ based on a two-dimensional closed loop stage actuator (Model no P517.2CL; Physik Instrumente (PI) GmbH & Co. KG, Germany). The whole sample consists of a 30-µl droplet of a mixture of 9.6 µm beads, bacteria and 3 µm beads. The drop is placed on a trichloroethylene treated (hydrophobic) coverslip attached to the bottom of the slide. The sample chamber is in turn sealed with a second coverslip placed on the other side of the aluminum slide. The use of a sealed slide minimizes the evaporation from the sample and thereby any undesired flow. Moreover the use of cover glass on top of the sample facilitates the use of an oil immersion condenser, which is important for precise imaging of the object onto the position sensitive device, (PSD, model no. L20 SU9; Sitek Electro Optics, Sweden). The collected probe laser light is deflected from the illumination path onto the PSD, via a polarizing beam splitter plate (model no. 03BTL025; Melles Griot, Täby, Sweden). Moreover, any remaining laser light from the trapping laser is filtered out with a laser line filter, LL (model no. XLK12; Omega Optical Inc, Brattleboro, VT, USA).

**2.2 Position detection system**

Position detection is carried out by focusing the weaker HeNe laser a short distance below the trapping focus, as is schematically illustrated in *Fig. 3* below. As a result, an enlarged light spot of the trapped bead appears in the far field. The position of this spot is a sensitive measure of the position of the trapped object.

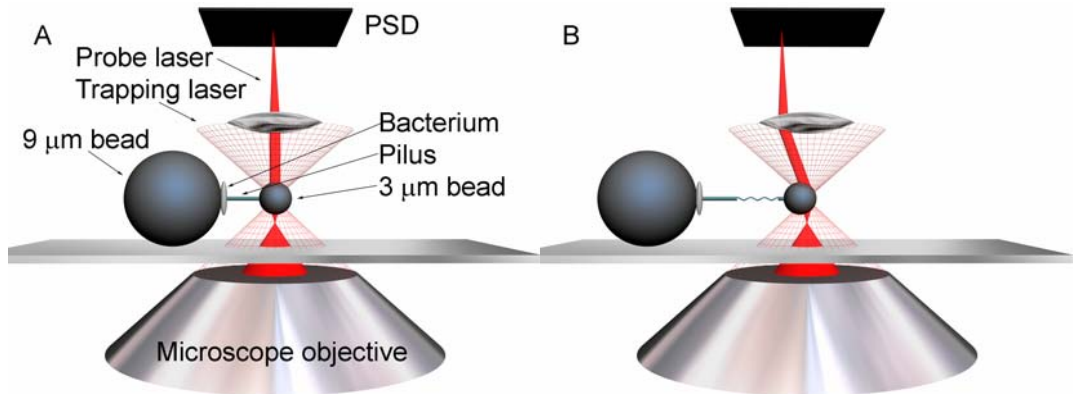

*Figure 3. A schematic illustration of the probe beam and the trapping beam for a bead positioned at the center of the trap and a bead that is slightly shifted off center. A and B illustrate the system when the pili is un-stretched and when the pili is put under tension, respectively.*

## 3. THE FORCE MEASUREMENT SYSTEM

The force measurement system for measurement of adhesion forces in biological applications is computer controlled based on LabVIEW. A fully automated system serves for repeatable measurements so as to eliminate any user dependent errors. The force measurement system is based on four automated procedures used frequently; 1) **Probe optimization**; a routine for optimization of the z-position of the probe laser. 2) **Linear response**; a routine for calibration of the normalized detector signal versus position in the trap, S [m$^{-1}$]. 3) **Power spectra**; a routine that derives the stiffness (force constant) of the trap, k. 4) **Measurement**; a number of routines for automatic force measurements.

In addition to the above mentioned automated procedures, two other routines that are of importance for accurate measurement of forces have been incorporated into the system. They only need to be used if an optical component in the system has been altered or for validation purposes. The first one is for **calibration of the trap response**, h[μm/μrad], i.e. how far the trap is moved when the GMM is tilted a certain angle. The other is a routine for **validation of the calibration of the force constant assessed by the power spectra routine** by the viscous drag force technique.

**3.1 Calibration of the trap response**

When the force measurement system has been properly aligned calibration of the trap response has to be performed. The trap response depends on the optical system between the GMM and the microscope objective whereby it normally only has to be performed once for a particular system. The calibration of the trap response is performed by the use of a piezo controlled mirror, a piezo stage and a high resolution camera. First the resolution of the imaging system has to be assessed, i.e. its dpμ (dots per micrometer). The dpμ of the system is derived through two pictures of a sample displaced a precise distance by the piezo stage (*Fig. 4 A and B*) The trap response can then be found by a set of similar pictures of a trapped object, created by a tilting of the gimbal mounted mirror a certain angle *(Fig. 4 C and D)*. The trap response of our system was found to be 0.00207 μm/μrad.

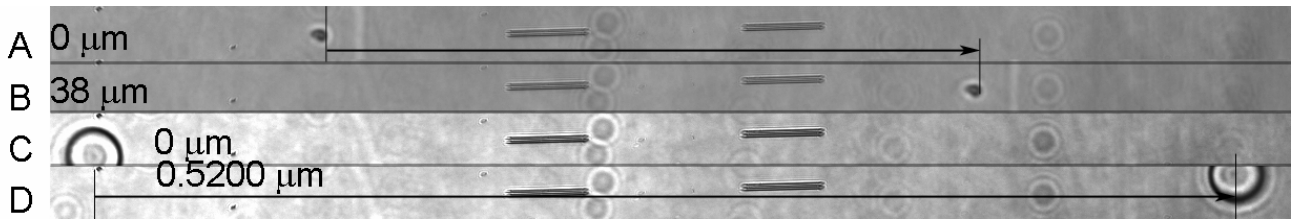

*Figure 4. By the use of the piezo stage and a high resolution camera it was possible to calibrate the angular response of the trapping position to a high accuracy. Panel A and B, the sample has been moved 38 μm by the piezo stage. Panel C and D, the bead was moved by the stepper motor that controls the tilt of GMM1 (0.52 μm). The angular response of the trapping position was found to be 0.00207 μm/μrad.*

### 3.2 Probe optimization

Probe optimization is a routine for control of the position of the probe laser relative to the trapping laser. The position of the focus of the probe laser is controlled by alternating the divergence of the probe laser beam, i.e. by changing the position of lens L4 via a stepper motor (*Fig. 1*). A non-linear response results when the trapping- and probing lasers are misaligned or when the bead is moved too far away from the equilibrium position. Cautious adjustments of the GMM1 mirror need therefore to be done until a linear signal is obtained. Typically, a bead with a diameter of 3 μm provides a linear detector signal if the displacement from the equilibrium position is less than 0.5 μm. Moreover, the focus of the probe laser relative the trapping laser is optimized so as to give highest sensitivity of the position measurements while achieving a system where the response of the force is insensitive to movement of the bead in the z-direction. The curves in *Fig. 5* correspond to a set of probe laser positions, obtained after running the probe optimization routine.

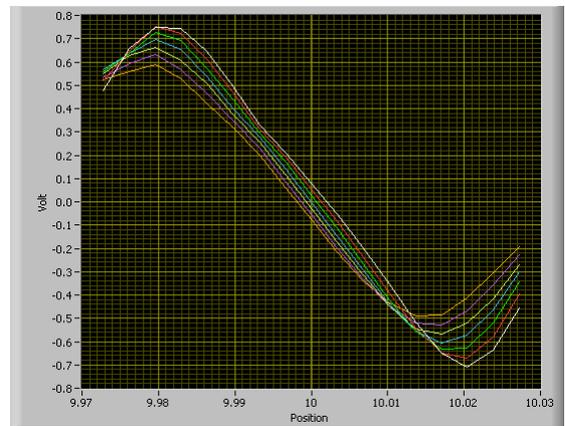

*Figure 5. The detector response for 7 different probe laser positions for a bead of diameter 3 μm as function of tilting of GMM1. The linear region extends to ± 0.5 μm.*

### 3.3 Linear response

Whence the probe laser beam is optimized the linear response S of the system is measured, i.e. the relation between the detector signal and the displacement of the bead. The displacement of the bead is chosen to be within the linear part of the detector response found in *Fig. 5*. The calibration is performed over a displacement of ± 0.4 μm. This calibration is performed before every individual measurement. *Fig. 6* shows a typical detector response for a three micrometer bead. The linear response is achieved by displacing the trapped object a small distance by a slight tilt of the piezo controlled gimbal mounted mirror. The position–signal response is in turn found by linear regression, see *Fig. 6*.

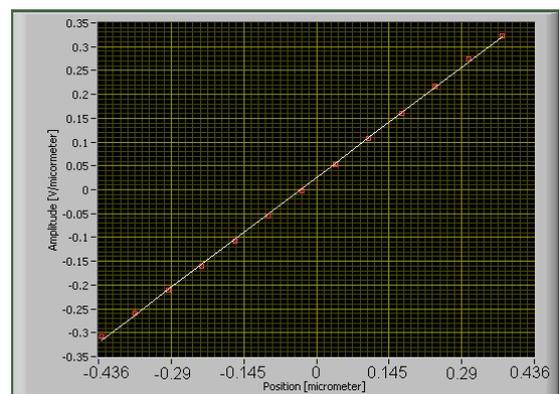

*Figure 6. The detector response for a bead with a diameter of 3 μm as function of tilting of GMM1. The linear region extends to ± 0.4 μm. This particular measurement gave a slope of 0.79* V/μm

## 3.4 Power spectrum

Several methods exist to correlate the measured position values to the applied force. A review of different methods is given in Visscher *et al.* [18]. One method to measure the force constant (or stiffness) of the trap, $k$, is to use the power spectrum, based on the thermal fluctuations (Brownian motion) of the trapped particle. The power spectrum of the Brownian motion, $S_x(f)$, can be expressed as

$$S_x(f) = |X(f)|^2 = \frac{k_B T}{\gamma \pi^2 \left( f_c^2 + f^2 \right)} \tag{1}$$

where $f_c = k/2\pi\gamma$ is the characteristic or break frequency, $k_B$ Boltzmann's constant, and $T$ the absolute temperature. The two latter factors originate from the power spectrum of the thermal force [18]. An example of a power spectrum of the Brownian motion of a trapped 3 µm spherical bead is shown in *Fig. 7*. At high frequencies, i.e. for $f \gg f_c$, $S_x(f)$ falls off like $1/f^2$ (the slope is –2 in the logarithmic diagram in *Fig. 7*). This is characteristic of free diffusion. Due to the trapping, the low frequencies are suppressed, and the power spectrum becomes approximately constant for $f \ll f_c$. The force constant in this example could thereby be determined to k = 0.91 pN/nm.

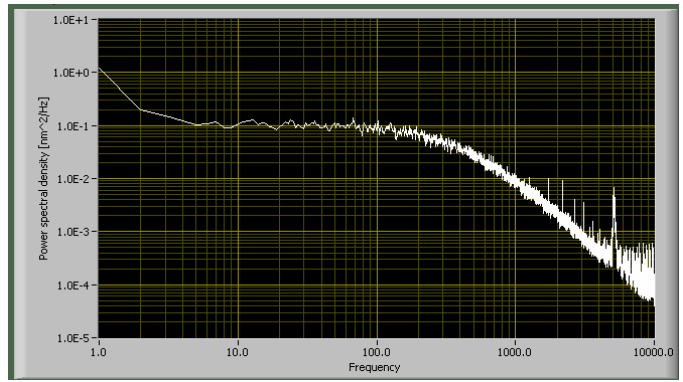

*Figure 7. Power spectrum of a 3 µm diameter bead moving in the optical trap. The break frequency, $f_c$ was found to be 321 Hz and the force constant, k, 0.91 pN/nm. The detector response was 0.79 V/µm whereas the laser power was 700 mW.*

## 3.4 Measurement

Whence the system has been properly calibrated, the force measurements are performed by the use of one of several automated measurement procedures that control the movement of the piezo-stage and collection of position data. All procedures are connected to a database where the measurement data and important information are stored. *Fig. 8* shows the front panel for measurement of adhesion forces. A variety of parameters, e.g. loading force, activation time, and retraction speed, can be set for investigation of adhesion mechanism.

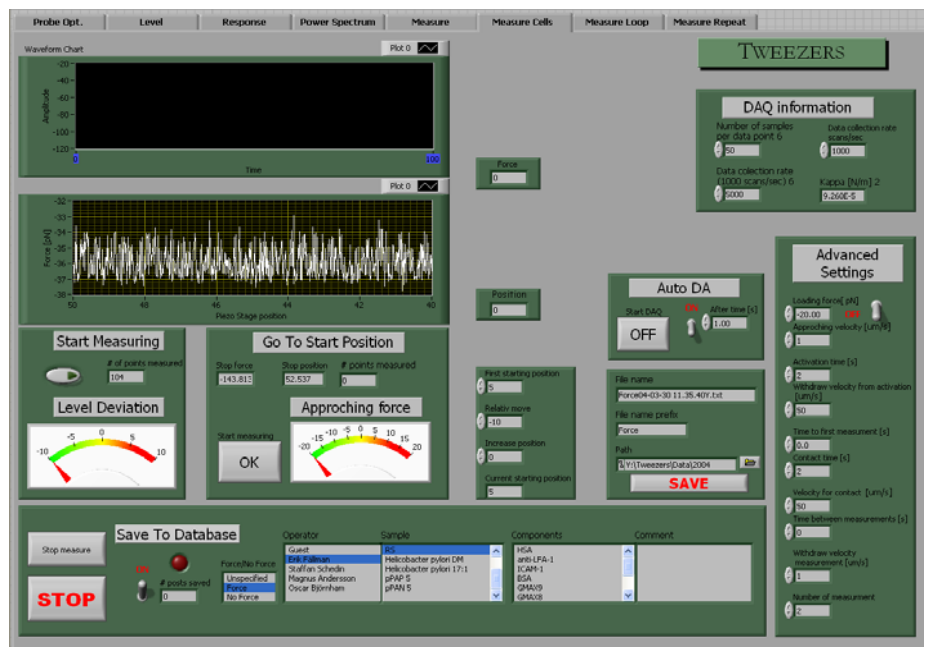

*Figure 8. The front panel of the force measurement system for measurements of adhesion forces. Parameters that can be set are: Number of repeated measurements, retraction speed, loading force, etc. The two graphs show the force-vs.-position and force-vs.-elongation data.*

## 4. MODEL SYSTEM AND EXPERIMENT

The force measurement system described above has for example been used for investigation of the mechanical properties of pili structures that mediates the binding of individual bacterium to host tissue. *Panel A of Figure 9* shows an AFM picture of an uropathogenic *Escherichia coli* expressing P pili (HB101/Ppap5) whereas panel B shows a schematic illustration of a pilus and its helical structure. A typical bacterium is ~half a micrometer wide and ~2 µm long and expresses pili structures with an average length of ~2 µm. In order to investigate the physical properties of a single pilus structure a biological model system and a measurement procedure have been developed. *Figure 3* above illustrates the model system. *E. coli* expressing P pili were covalently attached to large polystyrene beads (9.6 µm diameter) by an NHS linker. The large beads were immobilized onto a hydrophobic cover glass surface[16]. A second smaller polystyrene bead served as a handle for the optical tweezers. The small bead (3 µm diameter), functionalized with galabiose, bond both specifically and non-specifically to the P pili adhesin and the PapA rod, respectively. The entire system was contained in phosphate buffered saline (PBS, with a pH of 7.5).

In practice, a bacterium was trapped by the optical tweezers at low power (~100mW at the laser) and brought into contact with the large bead in a way such that attachment occurred. A galabiose bead was trapped by the tweezers and brought to a position close to the bacterium at the height where the force measurement was to be performed The laser power was increased to ~700 mW at the laser and the calibration procedure was performed. Next, the galabiose bead was brought to a starting position close but not in contact with the bacterium so that only the pili bind to the bead, so as to avoid any other bacterial surface – bead interactions

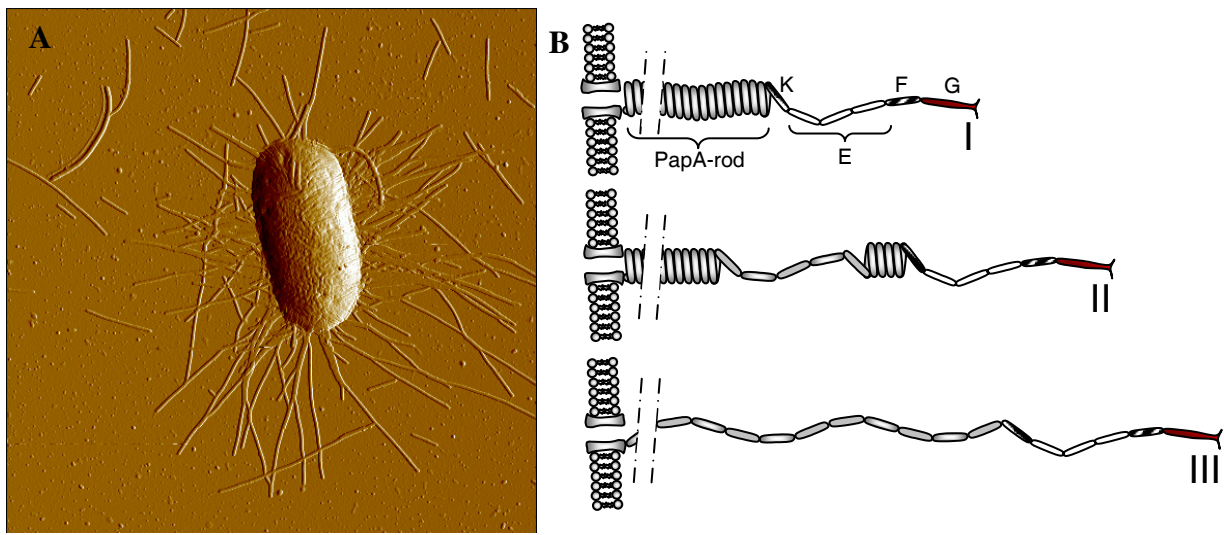

*Figure 9. Panel A shows an AMF image of a HB101/Ppap5 bacterium. The full image is 5 by 5 micrometer. Panel B shows a schematic illustration of a pilus under various degrees of elongation. I) A pilus that is stretched in its first elongation region (I) II) A pilus that is being unfolded (elongated in its second region). III) A pilus that is completely unfolded (elongated in its third elongation region).*

The measurement was initiated by moving the large bead away from the trapped galabiose bead at a controlled constant speed of 0.01-3 μm/s (depending on the experiment performed). The coverslip with the entire sample was moved by the piezo translator at an accuracy of 2 nm. When pili bind to the galabiose bead, held by the tweezers, a force is exerted on the binding as the bacterium is retracted from the trap. As a result, the galabiose bead is displaced a small amount from the center of the trap, which is a continuous measure of the binding force during retraction of the bacterium from the galabiose bead.

The force measurement system has been used for measurement of a number of physical/mechanical properties of P pili. We have, among other things, characterized three characteristic elongation regions of a single pilus, illustrated by I, II, and III in *panel B in Fig. 9* [14]. First, the pilus elongates elastically up to 15 – 20 % under an increased force, denoted region I (*Fig. 9 panel B, I*). At a force of ~27 pN the pilus enters region II where the helical structure of the PapA-rod unfolds under constant force (*Fig. 9 panel B, II*). The unfolding elongates the pilus to a length of ~700% of its original length. When the pilus is completely unfolded it enters its region III, referred to as the "s"-shaped region because of the form of its force-versus-elongation behavior (*Fig. 9 panel B, III*), which is believed to correspond to an overstretching of the unfolded PapA structure.

A typical force response for a HB101/Ppap5 bacterium is shown in *Fig.10*. In this particular case, the force was mediated by several pili up to 1.4 μm, characterized by a complex response consisting of several linear regions separated by a number of kinks (representing the detachment of pili). After 1.4 μm all pili except one detached from the small bead and the force was mediated by only one pilus, which already was in region II. The remaining pilus elongated to a length of 4 μm at which the helical structure of the PapA rod had been completely unfolded and entered region III. At 6.3 μm the remaining pilus detached and the galabiose bead retracted to the center of the trap, indicated by a zero force reading.

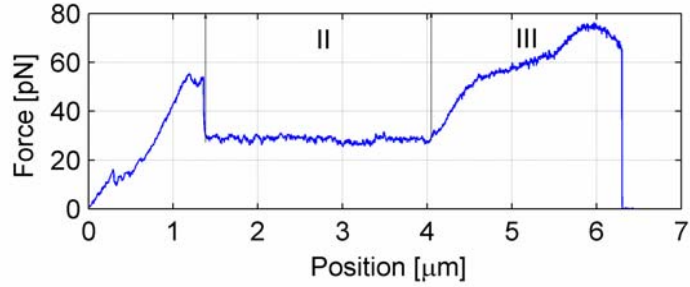

*Figure 10 A typical response from a force measurement on a bacterium carrying P pili, HB101/Ppap5. The parts of the force curve where the force is mediated by a single pilus are marked with appropriate region.*

The characteristic force-versus-elongation response of a single pilus $F_p$ can be modeled through the mathematical expression given in Equation (1) [14]

$$F_p(x) = \begin{cases} 0 & x \leq x_0 & \text{Before streching} \\ F_I(x) = k_p(x - x_0) & x_0 \leq x \leq x_1 & \text{Region I} \\ F_{II} = F_{uf} & x_1 \leq x \leq x_2 & \text{Region II} \\ F_{III}(x) & x_2 \leq x & \text{Region III} \end{cases} \quad (1)$$

where $x$ represents the distance between the bacterium and the surface to which the pilus bind (the small bead) and $x_0$ the unstretched length of the part of the pilus that mediates the binding (which is equal to the distance between the bacterium and the trapped bead when the binding starts to support forces, see *Fig. 3*). $x_1$ and $x_2$ are the bacterium-bead distances at which the pilus passes from region I to region II, and from region II to region III, respectively.

The first elongation region is for a single pilus seldom observed in force measurements because the initial response is almost always mediated by several pili. The regions II and III can, however, often be identified, e.g. as is displayed in *Fig. 10*. In addition to studies of the elongation properties of P pili, also contraction properties have been investigated. These are done by simply reversing the direction of the movement of the cover glass at a given elongation. Since this can be done for any elongation, it is possible to study the retraction behavior a single pilus by reversing the direction of the cover glass when the binding is mediated by solely one pilus. One such example is shown in *Fig. 11*. This response originates from the retraction of a single pilus elongated into its region III. The response shows that the elongation of the pilus is reversible [15], and thus not plastic in any of its unfolding regions, which previously has been assumed [10].

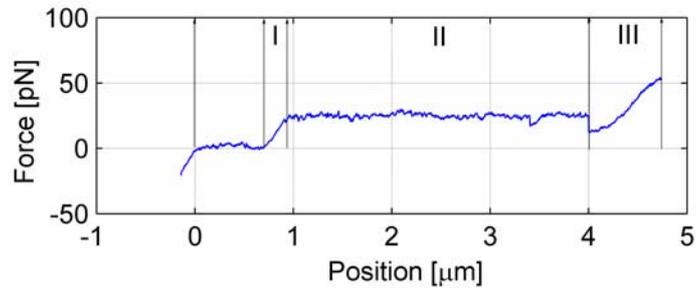

*Figure 11. A typical force measurement of the forces mediated by a single pilus during retraction. The three elongation regions are indicated. A retraction measurement gives at low contraction speeds a response that is identical to that of an unfolding measurement except for the small kink at the beginning of region three originating from a contraction of a linear conformation of the PapA rod [15].*

Ongoing investigations are concerned with the repeatability of the unfolding-folding process as well as its dependence on the contraction speed. It has been found that the folding of the helical structure of the PapA rod has a time-dependence. For low contraction velocities (~ 0.01 μm/s) the folding force is equal to the unfolding force (as can be seen by comparison of the *Figs. 10 and 11*. For higher retraction velocities, however, the retraction force drops occasionally down to a lower level, most often ~12 pN. This is interpreted as a temporary miss-folding. The repeatability is however good. For lower contraction velocities, contraction and retraction have large similarities. Also for higher retraction velocities the repeatability is good, although the miss-folding typically occur at dissimilar positions.

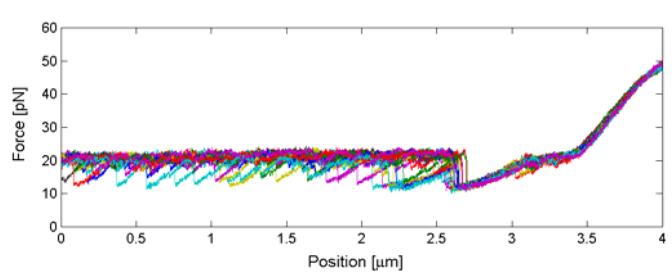

*Figure 12. The contraction force measured for 30 successive complete unfoldings of a PapA rod of a single P pilus.*

*Figure 12* displays a measurement where a single pilus has been unfolded and refolded 30 times. Each curve in the graph displays the refolding process after an unfolding event. The refolding speed was set to 0.2 μm/s. This measurement clearly indicates that a single pilus can undergo a significant number of unfolding-refolding cycles without loosing its intrinsic capability to retain its original shape after unfolding due to external stress, e.g. due to the rinsing force caused by increased urine flow.

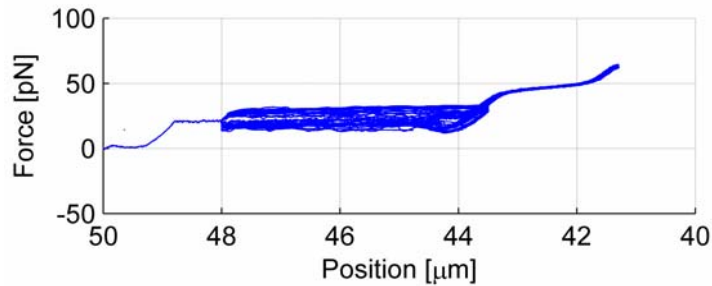

*Figure 13. Measurement data for the un-folding and folding force for a number of different loading rates, i.e. speed of the retraction. The measurement was done under increasing speed, from 0.2 µm/s up to a speed of 3 µm/s.*

Moreover, preliminary results from dynamic measurements shows that the unfolding force increases and the refolding force decreases when the speed is increased (*Fig. 13*). The origin to this behavior is under investigation.

## 5. CONCLUSIONS

A versatile force measurement system based on optical tweezers has been presented. The system allows time-resolved force measurements on the pN-level between micrometer-sized biological objects. Results of real-time force measurements of single P pilus on *E. coli* demonstrate the performance of the system. P pili from *E coli* have been characterized with respect to their elongation and retraction properties under exposure to external forces. The unfolding force of the helical (quaternary) structure of the PapA rod of P pili has been determined. Contraction studies have shown that the unfolding of P pilus is reversible, and not plastic, as previously has been assumed [10]. Studies of the repeatability and velocity dependence of the folding are under way.

In some applications it may be desirable to study the deformation between objects due to a constant force, rather than measuring a time-varying force. Such measurements can be realized in our system by feedback control of the movement of the piezo translator stage, such that the displacement from the equilibrium position of the trapped bead is held constant. As a result, a constant force will be exerted between the objects under study.

## 6. ACKNOWLEDGEMENTS

This project was supported by the Swedish Research Council (Vetenskapsrådet).